\documentclass{elsart}
\usepackage{epsfig,amssymb}

\journal{Physics Letters B}
\begin{document}
\begin{frontmatter}

\title{
Study of $\eta'\rightarrow\eta\pi^+\pi^-$ Dalitz plot.
}

\author[IHEP]{V.Dorofeev},
\author[IHEP]{R.Dzheliadin},
\author[IHEP]{A.Ekimov},
\author[IHEP]{A.Fenyuk},
\author[IHEP]{Yu.Gavrilov},
\author[IHEP]{Yu.Gouz},
\author[IHEP]{A.Ivashin},
\author[IHEP]{V.Kabachenko},
\author[IHEP]{I.Kachaev},
\author[IHEP]{A.Karyukhin},
\author[IHEP]{Yu.Khokhlov},
\author[IHEP]{A.Konoplyannikov},
\author[IHEP]{V.Konstantinov},
\author[IHEP,MITP]{M.Makouski},
\author[IHEP]{V.Matveev},
\author[IHEP]{V.Nikolaenko},
\author[IHEP]{A.Ostankov},
\author[IHEP]{B.Polyakov},
\author[IHEP]{D.Ryabchikov},
\author[IHEP,MITP]{S.Shushkevich},
\author[IHEP]{A.A.Solodkov},
\author[IHEP]{A.V.Solodkov},
\author[IHEP]{O.Solovianov},
\author[IHEP]{E.Starchenko},
\author[IHEP]{A.Zaitsev},
\author[IHEP]{A.Zenin }
{
\address[IHEP]{IHEP, 142281, Protvino, Moscow region, Russia}
\address[MITP]{Moscow Institute of Physics and Technology,
141700 Dolgoprudny, Moscow region, Russia}
}

\begin{abstract}
Dalitz plot of the $\eta' \rightarrow \eta\pi^+\pi^-$ decay
is studied using the data collected with the VES spectrometer
in two different exclusive reactions. The coefficients 
for the matrix element squared decomposition are measured on the 
largest statistics of $\eta'$ decays reported so far.

\end{abstract}
\end{frontmatter}


\section{Introduction.}

%

 In spite of the long history, the $\eta'\to\eta\pi\pi$ decay is still not well
studied and comprehended. The interest in the given decay
is stimulated primarily by the nature of $\eta'$-meson itself --- the large
gluon component of $\eta'$-meson should have strong influence on the dynamics
of its dominant decay $\eta'\to\eta\pi\pi$~\cite{ref:novikov}.

   Theoretical treatment of the $\eta'$ decays
is based as a rule on the effective chiral Lagrangians (ChPT).
Early calculations of the 
$\eta' \rightarrow \eta\pi^+\pi^-$
decay matrix element (within the lowest order of the expansion)
resulted in the width estimation which was few times 
less than the experimental value (see ref.  \cite{ref:width}).
The analogous calculations in paper \cite{ref:slope} for the Dalitz-plot 
parameters have
overestimated the slope significantly.
Phenomenologicaly the mass of the
$\eta'$ is comparable with the theory scale and this can
degrade the accuracy of the momentum expansion.

The rescattering in the final state could lead to significant 
contribution of the meson
resonances, in particular of 
the scalar octet, up to their dominance ("sa\-tu\-ration")
in the coupling constants (see \cite{ref:reson} and references therein).
The high mass of the $\eta'$ makes its decays sensitive to the parameters
of the intermediate resonances in the relevant approaches.

   The Dalitz plot for decay $\eta'\rightarrow\eta\pi^+\pi^-$
is usually expressed in variables \\
\begin{equation}
X=\frac{\sqrt 3 }{Q} (T_{\pi^+}-T_{\pi^-}); \:\:\:\:
Y=\frac{m_\eta + 2m_\pi}{m_\pi } \frac{T_\eta}{Q} - 1;
\end{equation}
here $T_\eta$, $T_\pi$ denote the kinetic energies of mesons in
the $\eta'$ rest frame, and $Q = T_\eta + T_{\pi^+} + T_{\pi^-}$
\footnote{The reconstructed $\eta'$ mass and the kinetic energies used,
the systematics from experimental smearing should be corrected 
by Monte Carlo methods.}.
There are two different parametrizations of event density on Dalitz plot
in literature: the PDG \cite{ref:PDG} parametrization (which is also called linear
parametrization)
\begin{equation}
|M|^2 = | 1 + \alpha Y|^2 + cX + dX^2 ;
\end{equation}
and more general decomposition
\begin{equation}
|M|^2 = 1 + aY + bY^2 + cX + dX^2 ;
\end{equation}
where $\alpha$ is a complex parameter and $a,b,c,d$ are real parameters.
Parameter $c=0$ if the Charge Parity conservation holds.
Parametrizations (2) and (3) are equivalent if $b>\frac{a^2}{4}$. 
A measurement of parameter $\alpha$ 
at the statistics of $\sim 1400$ events
is reported in publication  \cite{ref:KALBFLEISCH74}, here the $Im(\alpha)$ is neglected,
and $Re(\alpha)=-0.08\pm0.03$. Other measurement is performed
by CLEO collaboration \cite{ref:CLEO} at the statistics of
$\sim 6700$ events and significant background, it yields 
$Re(\alpha)=-0.021\pm0.025$.  

   Matrix element of the $\eta'\rightarrow\eta\pi^0\pi^0$ decay
has been studied in \cite{ref:GAMS}
at the statistics of 5400 decays, the value $Re(\alpha)=-0.058\pm 0.013$
has been obtained.

   Decay channel $\eta'\rightarrow\eta\pi^+\pi^-$ 
has been used already  in  search for C-parity violation
\cite{ref:RITTENBERG}.
No statistically significant asymmetry between $\pi^+$ and $\pi^-$
has been observed, the measured asymmetry is $\:-0.04\pm0.08$.

   Measured form-factors of the Dalitz plot have been used 
in a study of hypothetical  q\~q  and  qq\~q\~q  nonets of scalar mesons
in bag model framework \cite{ref:Intemann}.
Later on the $\eta'$ decay width and 
form-factors have been considered in the chiral Lagrangian approach 
\cite{ref:Deshpande,ref:Schechter,ref:Beisert,ref:Borasoy,ref:Borasoy200510}. 
Measured values of parameter $\alpha$ in $\eta'\rightarrow \eta\pi^0\pi^0$
and $\eta'\rightarrow \eta\pi^+\pi^-$ decays are used in a study 
of possible nonet of scalar mesons
in  article \cite{ref:Schechter}.
Values of parameters $a,b,d$ in the parametrization (3) are
discussed in ChPT, taking into account
the final state interactions  
in paper \cite{ref:Beisert}.
This approach is developed 
in recent reports \cite{ref:Borasoy,ref:Borasoy200510}. 

 The present work extends our previous study
\cite{ref:RYABCH}.

\begin{figure}
  \includegraphics[height=.3\textheight]{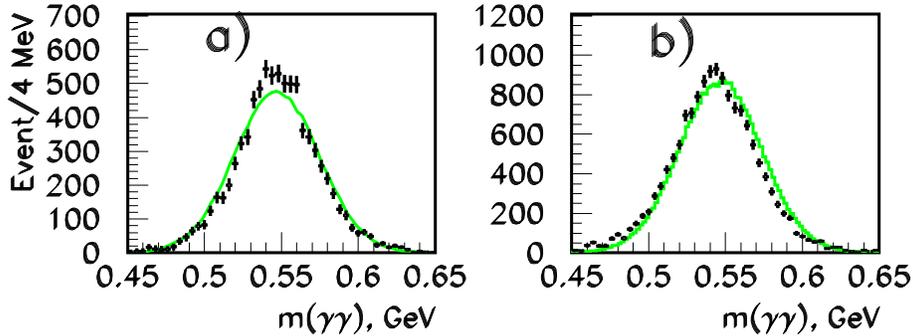}
  \caption{
Reconstructed $\eta$-mass distributions (Real data, histograms
 and Monte Carlo, smooth ~curves)~~
a) in Diffractive-like reaction;
b) in charge-exchange reaction.
Events from the $\eta'$ peak are selected. Background under 
$\eta'$ is subtracted.
}
\end{figure}

\section{Experimental procedure.}
   This study is based on the statistics acquired by VES
experiment for interactions of $\pi^-$ beam at 
the momentum of 27 $GeV/c$ on $Be$
target, in reactions $\pi^-p\rightarrow\eta'n$ 
(charge-exchange, $\sim 14.6\cdot10^3$ decays above background accepted for 
the analysis)
and $\pi^-N\rightarrow\eta'\pi^-N$ (diffractive-like production,
$\sim 7\cdot10^3$ decays).
The $\eta$ mesons off $\eta'$ decays were detected in 
$\eta\rightarrow\gamma\gamma$  mode
\footnote{Decay mode
$\eta\rightarrow\pi^+\pi^-\pi^0$ is not well suitable for this study  because of
confusion between charged particles 
originating off $\eta'$
and $\eta$ decays.}.
The background level within the mass band of $\pm21$ MeV near the $\eta'$ mass
was estimated as $\sim 5\%$ for the charge-exchange channel and $\sim 15\%$
for the diffractive-like channel.
Standard background subtraction procedure was applied for both reactions.
A  central mass band of $\eta\pi\pi$ system near $\eta'$ mass was taken
and also two side bands. The contribution of events 
from the side bands into distributions
were subtracted from the contribution of the central band
\footnote{Mass bands of different width were tested, from 13 $MeV$ to
21 $MeV$ for side bands and the double size for the central band, respectively.
The central bands of $\:\pm 21$ MeV width 
for both reactions are indicated by 
lines in Fig.2}.

The setup description and
selection procedure for diffractive-like channel is described 
in \cite{ref:RYABCH}.
The following selection procedure was applied for the 
charge-exchange sample. 
Events with two charged tracks of opposite charge and two showers
in electromagnetic calorimeter were selected,
the total event energy was requested at the range from 
25 to 30 $GeV$.
Events with identified $e^-/e^+$ were excluded.
Both electromagnetic showers should have energy greater than 400 $MeV$, 
should not be  associated with charged tracks. 
Distributions of effective mass $m(\gamma\gamma)$ 
are shown in Fig.1.  
The effective mass
of the $\gamma\gamma$ pair should be between 455 and 620  
$MeV$.
For successful candidates, the reconstructed
momenta of two $\gamma$ were subjected to kinematical 1-C fit 
to the $\eta$ mass, and the fitted parameters of $\eta$-meson were
used in further analysis. Finally,  events with 
$0.9<m(\eta\pi^+\pi^-)<1.0\:\: GeV $ were selected.
Effective mass distribution of $(\eta\pi^+\pi^-)$
system is shown in 
Fig.2, together with corresponding Monte-Carlo distribution. 
The fitted gaussian width of 
the $\eta'$ peak is $\sim 7.4\:\: MeV$. 

\begin{figure}
  \includegraphics[height=.3\textheight]{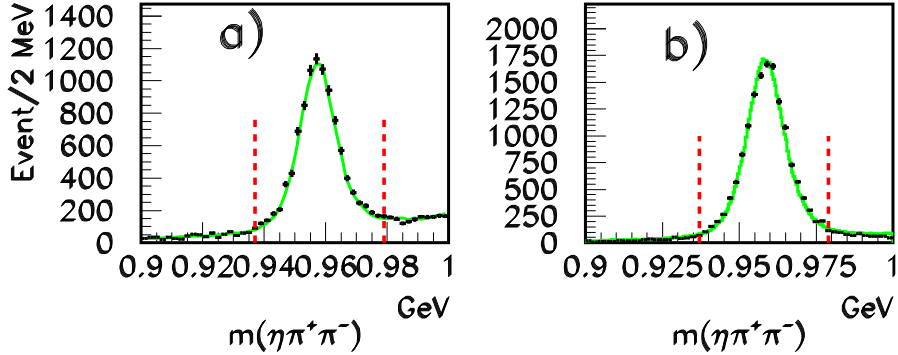}
  \caption{
Mass spectrum of $\eta\pi^+\pi^-$ sample produced in
a) diffractive-like reaction;
b) charge-exchange reaction. 
Points with errors and smooth curves represent the Real data and
MC respectively. Vertical lines  indicate the "signal mass band" for 
both samples.
}
\end{figure}
\begin{figure}
  \includegraphics[height=.30\textheight]{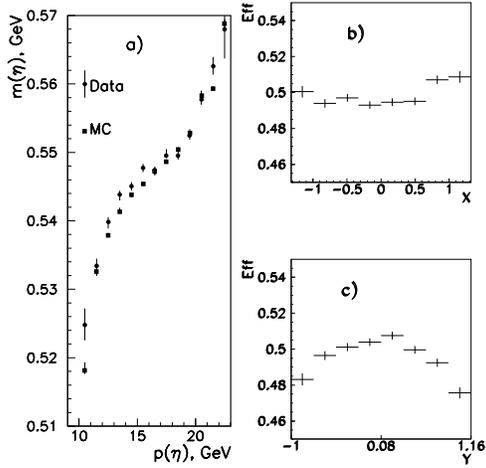}
  \caption{
Distributions for the charge-exchange sample:
a) correlation between measured mass of $\eta$ meson (before fit) and
its momentum in the laboratory frame for Real Data and Monte Carlo;
b) $\eta'$ detection efficiency as a function of X;
c) the efficiency as a function of Y variable.
}
\end{figure}

\begin{table}
\caption{Dalitz Plot  parameters for parametrization (3)
(Diffractive-like and charge-exchange data separately) }
\begin{tabular}{lcc}
\hline
                       Coeff. 
  &                      Diffractive-like   
  &                     Charge-exchange   \\
\hline
$a$ (at $Y$)       & $-0.093\pm 0.028 \pm 0.014(syst)$ & 
$-0.141  \pm 0.019 \pm 0.010(syst)$  \\
$b$ (at $Y^2$)     & $-0.129\pm 0.050 \pm 0.025(syst)$ & 
 $-0.096 \pm 0.033 \pm 0.016(syst)$  \\
$c$ (at $X$)       & $-0.015\pm 0.019 \pm 0.018(syst)$ & 
 $+0.028 \pm 0.013 \pm 0.020(syst)$  \\
$d$ (at $X^2$)     & $-0.128\pm 0.030 \pm 0.021(syst)$ & 
 $-0.062 \pm 0.021 \pm 0.009(syst)$ \\
\hline
${\chi^2/ND}$ &    69.5~/~55        & 
 52.1~/~55  \\
\hline
\end{tabular}
\end{table}

\begin{figure}
  \vbox to -1 cm {}
  \includegraphics[height=.45\textheight]{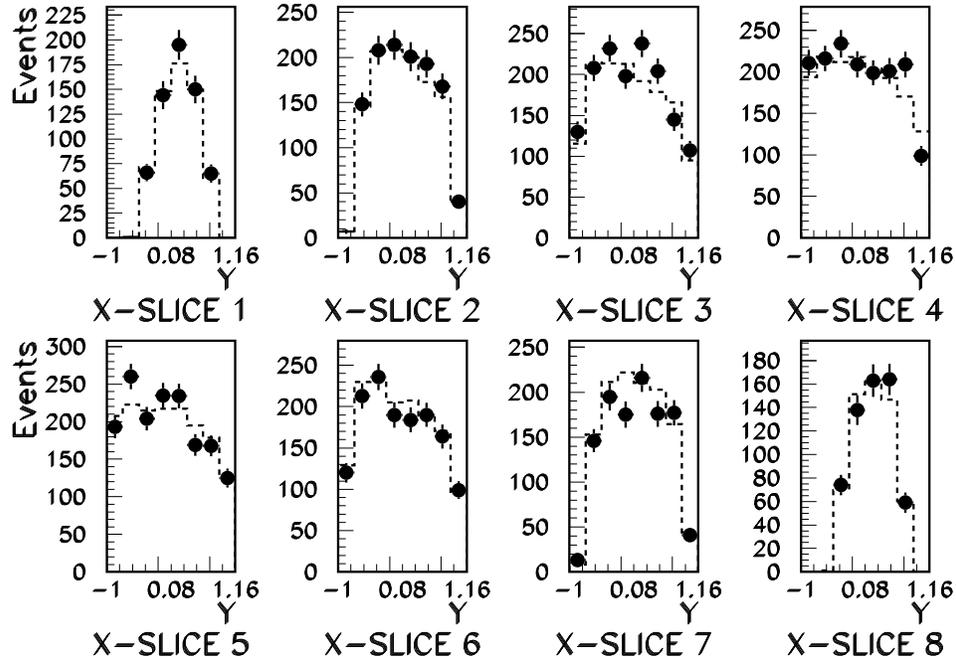}
  \caption{
Comparison ~of ~Real ~Data ~and ~Monte-Carlo ~weighted 
~with ~fitted ~coefficients
~as ~a ~function ~of ~Y, ~in ~different ~X-intervals
for charge-exchange reaction.
}
\end{figure}

\begin{table}
\caption{Dalitz Plot  parameters for parametrization (3) (combined fit)
and theoretical values.}
\begin{tabular}{lcccr}
\hline
                       Coeff. 
  &                     Combined fit   &
                      Theory \cite{ref:Beisert}   &
                      Theory \cite{ref:Borasoy}   &
                      Theory \cite{ref:Borasoy200510}   \\
\hline
$a$ (at $Y$)       &
$-0.127 \pm 0.016 \pm 0.008(syst)$ & -0.093                                & 
$-0.116 \pm 0.024$ & $-0.132 \pm 0.019$ \\
$b$ (at $Y^2$)     &
$-0.106 \pm 0.028 \pm 0.014(syst)$ & -0.059 & 
$ 0.000 \pm 0.019$ & $-0.108 \pm 0.033$ \\
$c$ (at $X$)       &
$+0.015 \pm 0.011 \pm 0.014(syst)$  &        & 
                   &                    \\
$d$ (at $X^2$)     &
$-0.082 \pm 0.017 \pm 0.008(syst)$  & -0.003 & 
$ 0.016 \pm 0.035$ & $-0.046 \pm 0.022$ \\
\hline
${\chi^2/ND}$ &
 129.3/114 &  & \\
\hline
\end{tabular}
\end{table}

   Available statistics of $\eta'$ decays in this experiment 
is sufficient for precise determination of the Dalitz plot parameters
(variations from the uniform distribution are expected at the level of 10\% ).
This objective
requires a careful Monte-Carlo (MC) simulation.
The MC  chain included the
event generation,
the detector simulation
based on GEANT3 package,
the trigger simulation
and the reconstruction of charged tracks.
A special fast Monte-Carlo procedure was  developed
for gammas. 
Event generation for the charge-exchange reaction included
the isotropic decay of the $\eta'$ on a background,
which rises linearly with $m(\eta\pi\pi)$,
and isotropic decay $\eta\rightarrow\gamma\gamma$.
For the diffractive-like reaction, 
the distribution on mass $M_{\eta'\pi^-}$ and
the polar and azimuthal angles of $\eta'$ in each bin on the 
$M_{\eta'\pi^-}$ were generated according to the results
of Partial Wave Analysis of $\eta'\pi^-$ system \cite{ref:DOROFEEV}.
Reconstructed MC sample was passed through the same selection procedure
as the real data sample.
MC statistic contains 450000 accepted events after cuts
for charge-exchange channel and 1000000 events for diffractive-like channel.
The quality of the MC simulation 
is demonstrated
in Fig.2, where the simulated mass spectra are superimposed on the Real Data.
Another comparison is presented in Fig.3a, where 
the correlation of measured $\eta$ mass with its laboratory
momentum is shown for Real Data and MC.
Good description of this correlation in Monte-Carlo provides 
low level of systematic errors associated with $m(\gamma\gamma)$ cut.
Variations of the $\eta'$ detection efficiency over Dalitz plot
are small, as one can judge from Fig.3b and 3c.
 
    The fitting procedure was arranged in the following way.
The Dalitz plot was subdivided
to 8 X-bins and 8 Y-bins, i.e. 64 cells in total.
The mostly populated cell contains $\sim 250$ events for the charge-exchange sample.
Dalitz plot  parameters were obtained
at the result of minimization of the functional:
\begin{equation}
\chi^2(N_1,N_2,a,b,c,d) = \sum_{r=1,2}\sum_{i}^{n_{bin}} {(D_{ri} - N_r M_{ri})^2
\over \sigma^2_{ri}}
\end{equation}

Here the index $r$  enumerates two reactions considered,
the index $i$ enumerates cells in Dalitz plot
\footnote{Four empty cells  outside the  Dalitz plot
boundaries are excluded.},
$N_1$ and $N_2$ are  normalization factors,
$a,b,c,d$ are the Dalitz plot parameters.
The $M_{ri}$ and $D_{ri}$ stand for the number of (weighted) entries in
the $i$-th bin of the two-dimensional histograms
in the Dalitz variables for
  {\bf M}onte-Carlo and for the background-subtracted {\bf d}ata respectively.
Statistical error $\sigma$ includes background subtraction
and MC statistical errors.
The MC histogram is obtained as follows:

\begin{equation}
M_{ri} = \sum_{j=1}^{N_{ev}^r} (1+aY_{rj}+
bY^2_{rj}+cX_{rj}+dX^2_{rj}),
\end{equation}
where $j$ enumerates the generated events and $X_{rj}$ and $Y_{rj}$
are the reconstructed values in the event $j$.

   Table 1 represents the obtained sets of parameters 
for both production reactions separately
\footnote{Values given for the diffractive-like channel
supersede the results of previous publication \cite{ref:RYABCH}.
The difference is caused by more developed Monte-Carlo model in this study. 
}.
The quality of 
Monte Carlo description of Real Data 
is demonstrated in Fig.4.
Our estimation of systematic errors includes variation of parameters
with the width of mass windows and  background tuning in MC.
Instability of the fitted parameters with respect to
the variation of charge tracks detection efficiency in the vicinity of
non-interacting beam tracks in the detector is small and it 
is also included into
systematic errors.  

   Table 2 contains the estimation of the Dalitz plot parameters
from combined fit taking charge-exchange and diffractive-like
channels together. 
The $\chi^2$ for combined fit is satisfactory,
and one can conclude that
\begin{itemize} 
\item
parameter $b$ in parametrization (3) is negative at $3.8$ stand. deviation level,
therefore the parametrization (2) is not acceptable;
\item
parameter $d$ is negative;
\item
parameter $c$ is consistent with zero with error of order of $1\%$.
\end{itemize}

   Theoretical predictions from different publications 
\cite{ref:Beisert,ref:Borasoy,ref:Borasoy200510} for parameters $a,b,d$ 
are also given
in Table 2 for comparison.

Combined fit with "linear parametrization" (2)
yields unsatisfactory ${\chi^2/ND}=170.5~/~114$ ratio
and cannot be accepted.

   Attempting to clarify the origin of observed Y- and
X-curvatures on the Dalitz plot, we
adopted  the procedure described in 
article \cite{ref:Schechter} which contains a general
Chiral Lagrangian for interactions of pseudoscalar mesons
and a hypo\-the\-ti\-cal nonet of light scalar mesons.
However, the Y-dependence of the event density in this
model is close to linear and cannot reproduce the observed
quadratic term at Y parameter.

\section{Conclusions}

The Dalitz plot parameters for decay $\eta'\rightarrow\eta\pi^+\pi^-$
in two production reactions
are measured. 
We conclude that
\begin{itemize}
\item
results obtained for $\eta'$-mesons produced in charge-exchange
and in diffractive-like reactions are consistent;
\item
the "linear" parametrization (2) gives unacceptable
${\chi^2/ND}$ 
which is connected with significantly negative value of the parameter $b$
at $Y^2$ term 
in more general parametrization (3);
\item
the value of the Charge Parity nonconservation parameter $c$ 
is consistent with zero;
\item
observed quadratic term at $Y$-variable is not reproduced
in Effective Chiral Lagrangian model 
supplemented by nonet of scalar mesons
\cite{ref:Schechter},
however it can be adopted in Chiral Unitarized model
with interactions in final state
\cite{ref:Beisert,ref:Borasoy200510}.
The dynamical nature of this term needs clarification.
\end{itemize}

%
%
\begin{ack}
 This work is supported in part by the Russian Foundation of Basic Research
grants RFBR 05-02-08082 
and RFBR 05-02-17664  and by Presidential grant \\
NSh 5911.2006.2. 
\end{ack}
%
\bibliographystyle{aipproc}   
\IfFileExists{\jobname.bbl}{}
 {\typeout{}
 }

\end{document}